\documentclass[5p]{elsarticle}  

\usepackage[cmex10]{amsmath}
\usepackage{amssymb}
\usepackage{graphicx}
\usepackage{auto-pst-pdf}
\usepackage{psfrag}
\usepackage{verbatim}
\usepackage{mathdots}

\usepackage{mathrsfs}
\usepackage{nameref}

\usepackage{lineno,hyperref}
\modulolinenumbers[5]
\usepackage{natbib}

\usepackage{mathtools}

\usepackage{enumitem}
\newcounter{muni}
\newenvironment{remunerate}
               {\begin{list}{{\upshape 
               \arabic{muni}.}}{\usecounter{muni}
                \setlength{\leftmargin}{0pt}
                \setlength{\itemindent}{25pt}}}{\end{list}}

\makeatletter
\newcommand{\labitem}[2]{%
\def\@itemlabel{#1}
\item
\def\@currentlabel{#1}\label{#2}}
\makeatother

\def\smallint{\begingroup\textstyle \int\endgroup}

\title{On the internal signature and minimal electric network realizations of reciprocal behaviors}

\author[thh22]{Timothy H.\ Hughes}\ead{t.h.hughes@exeter.ac.uk}    
\address[thh22]{College of Engineering, Mathematics and Physical Sciences, University of Exeter, Penryn Campus, Penryn, Cornwall, TR10 9EZ, UK}

\begin{document}

\newtheorem{thm}{Theorem} 
\newtheorem{lem}[thm]{Lemma} 
\newdefinition{rem}[thm]{Remark} 
\newtheorem{example}[thm]{Example}
\newdefinition{defn}[thm]{Definition} 
\newproof{pf}{Proof} 
\providecommand{\abs}[1]{\lvert#1\rvert}

\begin{abstract}                          
In a recent paper, it was shown that (i) any reciprocal system with a proper transfer function possesses a signature-symmetric realization in which each state has either even or odd parity; and (ii) any reciprocal and passive behavior can be realized as the driving-point behavior of an electric network comprising resistors, inductors, capacitors and transformers. These results extended classical results to include uncontrollable systems. In this paper, we establish new lower bounds on the number of states with even parity (capacitors) and odd parity (inductors) for reciprocal systems that need not be controllable.
\end{abstract}


\maketitle

\section{Introduction}
It is well known that any symmetric transfer function $H$ possesses a so-called signature symmetric realization in which the states are partitioned into a number with even parity ($n_{1}$) and a number with odd parity ($n_{2}$) \cite{JWDSP2, RBA}. Furthermore, the numbers $n_{1}$ and $n_{2}$ are related to the properties of Hankel, Sylvester, and Bezoutian matrices associated with $H$ (see \cite{BitAnd, PFSRTF, HugSmAI}). In \cite{HugSmAI}, a physical consequence of these results was provided, which related to electric networks comprising resistors, inductors, capacitors and transformers (RLCT networks). Specifically, it was shown that the number of capacitors (resp., inductors) is bounded below by $n_{1}$ (resp., $n_{2}$) in any RLCT network whose impedance is $H$.

In \cite{THRB, THTPLSNA, HugNa, camwb}, it was noted that there are many important RLCT networks that are uncontrollable. Moreover, there are RLCT networks which do not possess an impedance. Also, even if the impedance does exist, then it does not fully determine the driving-point behavior when the RLCT network is uncontrollable. In \cite[Definition 4]{THRB}, a definition of reciprocity was provided which does not assume the existence of a transfer function (of relevance to RLCT networks which do not possess an impedance), and it was shown that any RLCT network is reciprocal. That paper then developed a theory of reciprocal systems that extends the classical results on signature symmetric realizations to systems that need not be controllable (see, e.g., \cite[Theorem 8]{THRB}). In particular, it was shown that any behavior that is reciprocal and passive (in accordance with \cite[Definitions 4 and 11]{THRB}) can be realized as the driving-point behavior of an RLCT network \cite[Theorem 13]{THRB}.

The results in this paper establish new lower bounds on the number of states with even parity (capacitors) and odd parity (inductors) for reciprocal systems. These bounds coincide with the aforementioned known lower bounds when the system is controllable, but they are higher for uncontrollable systems. In Theorem \ref{thm:ssrsp}, we show that, for any signature symmetric realization of a reciprocal system, the number of states with even (resp., odd) parity is bounded below by the sum of the number of uncontrollable modes and the number of positive (resp., negative) eigenvalues of a Bezoutian matrix derived from the high order differential equations describing the system. Theorem \ref{thm:brlctc} shows that, for RLCT realizations of passive and reciprocal behaviors, both an inductor and a capacitor are required to realize each uncontrollable mode. These theorems prove that the realization provided in Theorem 8 (resp., Theorem 13) of \cite{THRB} is minimal in the sense that it contains the least possible numbers of states with even parity (resp., capacitors) and states with odd parity (resp., inductors). To prove these results, we use the concept of the extended Cauchy index of a real-rational function, first defined in \cite{HugSmAI}, and the related concept of the McMillan degree. In particular, we obtain new bounds concerning the McMillan degree and extended Cauchy index of functions of the form $S^{T}HS$ where $H$ is a symmetric real-rational function and $S$ is a real matrix (see Theorem \ref{thm:pnsr}).

\section{Notation and preliminaries}
We denote the real numbers by $\mathbb{R}$. The polynomials, rational functions, and proper (i.e., bounded at infinity) rational functions in the indeterminate $\xi$ with real coefficients are denoted $\mathbb{R}[\xi], \mathbb{R}(\xi)$, and $\mathbb{R}_{p}(\xi)$. The $m {\times} n$ matrices with entries from $\mathbb{R}$ (resp., $\mathbb{R}[\xi]$, $\mathbb{R}(\xi)$, $\mathbb{R}_{p}(\xi)$) are denoted $\mathbb{R}^{m \times n}$ (resp., $\mathbb{R}^{m \times n}[\xi]$, $\mathbb{R}^{m \times n}(\xi)$, $\mathbb{R}_{p}^{m \times n}(\xi)$), and $n$ is omitted if $n=1$. If $H \in \mathbb{R}^{m \times n}, \mathbb{R}^{m \times n}[\xi]$, or $\mathbb{R}^{m \times n}(\xi)$, then $H^{T}$ denotes its transpose, and if $H$ is nonsingular (i.e., $\det{(H)} \neq 0$) then $H^{-1}$ denotes its inverse. If $H \in \mathbb{R}^{m \times n}$, then $\text{\#col}(H)$ denotes the numbers of columns ($n$); $\text{\#row}(H)$ denotes the number of rows ($m$); $\text{rank}(H)$ denotes the dimension of its column space: $\lbrace \mathbf{z} \in \mathbb{R}^{m} \mid \exists \mathbf{y} \in \mathbb{R}^{n} \text{ with } \mathbf{z} = H\mathbf{y}\rbrace$; and $\text{nullity}(H)$ denotes the dimension of its nullspace: $\lbrace \mathbf{z} \in \mathbb{R}^{n} \mid H\mathbf{z} = 0\rbrace$. As is well known, $\text{rank}(H)$ is also the dimension of the column space of $H^{T}$; $\text{rank}(H) + \text{nullity}(H) = \text{\#col}(H)$; and $\text{rank}(H) + \text{nullity}(H^{T}) = \text{\#row}(H)$. If $M \in \mathbb{R}^{m \times m}$ is symmetric, then all of the eigenvalues of $M$ are real; $M > 0$ ($M \geq 0$) indicates that $M$ is positive (non-negative) definite; and $\pi(M)$ (resp., $\nu(M)$) denotes the number of strictly positive (resp., strictly negative) eigenvalues of $M$. A matrix $\Sigma \in \mathbb{R}^{n \times n}$ is called a signature matrix if it is diagonal and all of its diagonal entries are either $1$ or $-1$. We denote the block column and block diagonal matrices with entries $M_{1}, \ldots , M_{n}$ by $\text{col}(M_{1} \hspace{0.15cm} \cdots \hspace{0.15cm} M_{n})$ and $\text{diag}(M_{1} \hspace{0.15cm} \cdots \hspace{0.15cm} M_{n})$. 

If $H \in \mathbb{R}^{m \times n}(\xi)$, then $\delta(H)$ denotes its McMillan degree. If, in addition, $H$ is symmetric, then $\gamma(H)$ denotes its extended Cauchy index (see Definition \ref{def:eci}). If $R \in \mathbb{R}^{m \times n}[\xi]$ has full row rank, then $\Delta(R)$ denotes the maximal degree among all $m \times m$ determinants formed from columns of $R$, and $\text{normalrank}(R) \coloneqq \max_{\lambda \in \mathbb{C}}(\text{rank}(R(\lambda)))$.

We consider behaviors (systems) defined as the set of weak solutions (see \cite[Section 2.3.2]{JWIMTSC}) to a differential equation of the form:
\begin{equation}
\hspace*{-0.3cm} \mathcal{B} = \lbrace \mathbf{w} \in \mathcal{L}_{1}^{\text{loc}}\left(\mathbb{R}, \mathbb{R}^{q}\right) \mid R(\tfrac{d}{dt})\mathbf{w} {=} 0\rbrace, \hspace{0.1cm} R \in \mathbb{R}^{p \times q}[s], \label{eq:bd}
\end{equation}
where $\mathcal{L}_{1}^{\text{loc}}\left(\mathbb{R}, \mathbb{R}^{q}\right)$ denotes the ($q$-vector-valued) locally integrable functions \cite[Defns.\ 2.3.3, 2.3.4]{JWIMTSC}. $\mathcal{B}$ is called \emph{controllable} if, for any two trajectories $\mathbf{w}_{1}, \mathbf{w}_{2} \in \mathcal{B}$ and $t_{0} \in \mathbb{R}$, there exists $\mathbf{w} \in \mathcal{B}$ and $t_{1} \geq t_{0}$ such that $\mathbf{w}(t) = \mathbf{w}_{1}(t)$ for all $t \leq t_{0}$ and $\mathbf{w}(t) = \mathbf{w}_{2}(t)$ for all $t \geq t_{1}$ \cite[Definition 5.2.2]{JWIMTSC}. A particular focus is on state-space systems:
\begin{align}
&\hspace*{-0.3cm} \mathcal{B}_{s} = \lbrace (\mathbf{u}, \mathbf{y}, \mathbf{x}) {\in} \mathcal{L}_{1}^{\text{loc}}\left(\mathbb{R}, \mathbb{R}^{n}\right) {\times} \mathcal{L}_{1}^{\text{loc}}\left(\mathbb{R}, \mathbb{R}^{n}\right) {\times} \mathcal{L}_{1}^{\text{loc}}\left(\mathbb{R}, \mathbb{R}^{d}\right) \mid \nonumber\\
& \hspace{1.0cm} \tfrac{d\mathbf{x}}{dt} = A\mathbf{x} + B\mathbf{u} \text{ and } \mathbf{y} = C\mathbf{x} + D\mathbf{u} \rbrace, \nonumber \\
& A \in \mathbb{R}^{d \times d}, B \in \mathbb{R}^{d \times n}, C \in \mathbb{R}^{n \times d}, D \in \mathbb{R}^{n \times n}. \label{eq:bhssr}
\end{align}
Here, we call the pair $(A,B)$ \emph{controllable} if $\mathcal{B}_{s}$ is controllable; and we call the pair $(C,A)$ \emph{observable} if $(\mathbf{u}, \mathbf{y}, \mathbf{x}) \in \mathcal{B}_{s}$ and $(\mathbf{u}, \mathbf{y}, \hat{\mathbf{x}}) \in \mathcal{B}_{s}$ imply $\mathbf{x} = \hat{\mathbf{x}}$ \cite[Definition 5.3.2]{JWIMTSC}.

We denote the behavior obtained by eliminating the state-variable $\mathbf{x}$ from $\mathcal{B}_{s}$ by $\mathcal{B}_{s}^{(\mathbf{u},\mathbf{y})} \coloneqq \lbrace (\mathbf{u}, \mathbf{y}) \mid \exists \mathbf{x}$ such that $(\mathbf{u}, \mathbf{y}, \mathbf{x}) \in \mathcal{B}_{s}\rbrace$. It can be shown that $\hat{\mathcal{B}} = \mathcal{B}_{s}^{(\mathbf{u},\mathbf{y})}$ takes the form
\begin{align}
&\hspace*{-0.3cm} \hat{\mathcal{B}} {=} \lbrace (\mathbf{u}, \mathbf{y}) {\in} \mathcal{L}_{1}^{\text{loc}}\left(\mathbb{R}, \mathbb{R}^{n}\right) {\times} \mathcal{L}_{1}^{\text{loc}}\left(\mathbb{R}, \mathbb{R}^{n}\right) \mid \hat{P}(\tfrac{d}{dt})\mathbf{u} {=} \hat{Q}(\tfrac{d}{dt})\mathbf{y}\rbrace, \nonumber \\ 
&\hspace*{-0.3cm} \hat{P}, \hat{Q} \in \mathbb{R}^{n \times n}[\xi], \hat{Q} \text{ nonsingular }, \hat{Q}^{-1}\hat{P} \text{ proper}.\label{eq:bgsqd}
\end{align}

More generally, for any given $T_{1} \in \mathbb{R}^{p_{1} \times q}, \ldots , T_{n} \in \mathbb{R}^{p_{n} \times q}$ such that $\text{col}(T_{1} \hspace{0.15cm} \cdots \hspace{0.15cm} T_{n}) \in \mathbb{R}^{q \times q}$ is a nonsingular matrix, and integer $1 \leq m \leq n$, we denote the projection of $\mathcal{B}$ onto $T_{1}\mathbf{w}, \ldots , T_{m}\mathbf{w}$ by
\begin{multline*}
\hspace*{-0.4cm} \mathcal{B}^{(T_{1}\mathbf{w}, \ldots , T_{m}\mathbf{w})} = \lbrace (T_{1}\mathbf{w}, \ldots , T_{m}\mathbf{w}) \mid \exists (T_{m+1}\mathbf{w}, \ldots , T_{n}\mathbf{w}) \\ \text{such that } \mathbf{w} \in \mathcal{B}\rbrace.
\end{multline*}

\section{Reciprocity and minimality}
\label{sec:ctr}
This section contains the formal statement of our main results (Theorems \ref{thm:ssrsp} and \ref{thm:brlctc}), which extend classical results on the minimal realization of reciprocal behaviors (Lemmas \ref{thm:Gpbm} and \ref{lem:rlctclb}) to systems which need not be controllable.  Motivation for considering such uncontrollable behaviors was outlined in \cite{THRB}. There, it was noted that many important reciprocal physical systems are uncontrollable, for example, the famous Bott-Duffin electric networks. That paper established a theory of reciprocal systems that extended classical results on signature symmetric realizations of behaviors to include such uncontrollable systems. However, \cite{THRB} did not address the question of minimality, which is the focus of the current paper. 

An equally important topic is passivity of uncontrollable behaviors, which was considered in \cite{THTPLSNA}. Both \cite{THRB} and \cite{THTPLSNA} considered behaviors of the form:
\begin{align}
\hspace*{-0.3cm} \mathcal{B} &{=} \lbrace (\mathbf{i}, \mathbf{v}) {\in} \mathcal{L}_{1}^{\text{loc}}\left(\mathbb{R}, \mathbb{R}^{n}\right) {\times} \mathcal{L}_{1}^{\text{loc}}\left(\mathbb{R}, \mathbb{R}^{n}\right) \mid P(\tfrac{d}{dt})\mathbf{i} {=} Q(\tfrac{d}{dt})\mathbf{v}\rbrace, \nonumber \\ 
& \text{with } P, Q \in \mathbb{R}^{n \times n}[\xi], \text{normalrank}(\left[P \hspace{0.25cm} {-}Q\right]) = n.\label{eq:bgd}
\end{align}

In \cite{HUGIFAC}, it was shown that the driving-point behavior of any passive electrical circuit necessarily has the above form, where $\mathbf{i}$ denotes the driving-point currents and $\mathbf{v}$ the corresponding driving-point voltages. We note that the partitioning $(\mathbf{i}, \mathbf{v})$ need not be an input-output partition in the sense of \cite[Definition 3.3.1]{JWIMTSC}, as $Q$ need not be nonsingular, and if $Q$ is nonsingular then $Q^{-1}P$ need not be proper. In this general setting, the papers \cite{THRB, THTPLSNA} defined reciprocity and passivity as follows. 
\begin{defn}[Reciprocal system]
\label{def:rb}
Let $\mathcal{B}$ be as in (\ref{eq:bgd}). $\mathcal{B}$ is called \emph{reciprocal} if, whenever $(\mathbf{i}_{a}, \mathbf{v}_{a}), (\mathbf{i}_{b}, \mathbf{v}_{b}) \in \mathcal{B}$ have bounded support on the left, then $\smallint_{-\infty}^{\infty}{\mathbf{v}_{b}(\tau)^{T}\mathbf{i}_{a}(t-\tau)d\tau} = \smallint_{-\infty}^{\infty}{\mathbf{i}_{b}(\tau)^{T}\mathbf{v}_{a}(t-\tau)d\tau}$ for all $t \in \mathbb{R}$.
\end{defn}

\begin{defn}[Passive system]
\label{def:pb}
$\mathcal{B}$ in (\ref{eq:bgd}) is called \emph{passive} if, given any $(\mathbf{i}, \mathbf{v}) \in \mathcal{B}$ and any $t_{0} \in \mathbb{R}$, there exists a $K \in \mathbb{R}$ (dependent on $(\mathbf{i}, \mathbf{v})$ and $t_{0}$) such that, if $t_{1} \geq t_{0}$ and $(\tilde{\mathbf{i}}, \tilde{\mathbf{v}}) \in \mathcal{B}$ satisfies $(\tilde{\mathbf{i}}(t), \tilde{\mathbf{v}}(t)) = (\mathbf{i}(t), \mathbf{v}(t))$ for $t < t_{0}$, then $-\smallint_{t_{0}}^{t_{1}} \tilde{\mathbf{i}}^{T}(t)\tilde{\mathbf{v}}(t) dt < K$.
\end{defn}

It is shown in \cite{THRB} that $\hat{\mathcal{B}}$ in (\ref{eq:bgsqd}) is reciprocal if and only if the transfer function $\hat{Q}^{-1}\hat{P}$ is symmetric. The paper \cite{THRB} then provided the following extension of a classical result on the existence of signature symmetric realizations to systems that need not be controllable.
\begin{lem}
\label{lem:bhrulb}
Let $\hat{\mathcal{B}}$ be as in (\ref{eq:bgsqd}). Then the following are equivalent.
\begin{enumerate}
\item $\hat{\mathcal{B}}$ is reciprocal.
\item There exists $\mathcal{B}_{s}$ as in (\ref{eq:bhssr}) and a signature matrix $\Sigma_{i} \in \mathbb{R}^{d \times d}$ such that (i) $\hat{\mathcal{B}} = \mathcal{B}_{s}^{(\mathbf{u},\mathbf{y})}$; and (ii) $A\Sigma_{i} = \Sigma_{i}A^{T}$, $B = \Sigma_{i}C^{T}$, and $D = D^{T}$.
\end{enumerate}
\end{lem}

If $\mathcal{B}_{s}$ is as in Lemma \ref{lem:bhrulb}, then we say that the entries in $\mathbf{x}$ corresponding to $+1$ (resp., $-1$) entries in $\Sigma_{i}$ have even (resp., odd) parity, so the number of states of even (resp., odd) parity is equal to $\pi(\Sigma_{i})$ (resp., $\nu(\Sigma_{i})$). 

Moreover, from \cite{THRB, HUGIFAC, THTPLSNA}, the driving-point behavior of any given RLCT network is necessarily passive and reciprocal, and we have the following result on the realization of passive and reciprocal behaviors.
\begin{lem}
\label{lem:brulb}
Let $\mathcal{B}$ in (\ref{eq:bgd}) be passive and reciprocal. Then there exists an RLCT network whose driving-point behavior is $\mathcal{B}$.
\end{lem}

If $\hat{\mathcal{B}}$ in Lemma \ref{lem:bhrulb} is controllable, then it can be shown that the number of states with even and odd parity are related to the properties of Hankel, Sylvester, and Bezoutian matrices associated with the polynomial matrices $\hat{P}$ and $\hat{Q}$ in (\ref{eq:bgsqd}) (see \cite{BitAnd, HugSmAI}). The most relevant results for this paper involve the Bezoutian matrix $\textnormal{Bez}(\hat{Q},\hat{P})$ defined as follows. 
\begin{defn}
\label{def:bezd}
Let $P, Q \in \mathbb{R}^{n \times n}[\xi]$, and let $m$ be the maximum of the degrees of the entries in $P$ and $Q$. Then $\textnormal{Bez}(Q,P)$ is the block matrix whose block entries $(\textnormal{Bez})_{ij}$ satisfy
\begin{equation*}
\frac{Q(z)P(w)^{T} {-} P(z)Q(w)^{T}}{z{-}w} = \sum_{i=1}^{m}{\sum_{j=1}^{m}{(\textnormal{Bez})_{ij}z^{i{-}1}w^{j{-}1}}}.
\end{equation*}
\end{defn}

It is easily verified that, for any given $P,Q \in \mathbb{R}^{n \times n}[\xi]$, $\textnormal{Bez}(Q,P)$ is uniquely specified by Definition \ref{def:bezd}. Also, if $Q$ is nonsingular with $Q^{-1}P$ symmetric, then $\textnormal{Bez}(Q,P)$ is symmetric. Then, from results in \cite{BitAnd, HugSmAI, THRB}, we have the following lemma.
\begin{lem}
\label{thm:Gpbm}
Let $\hat{\mathcal{B}}$ in (\ref{eq:bgsqd}) be controllable, and let $\mathcal{B}_{s}$ in (\ref{eq:bhssr}) be such that (i) $\hat{\mathcal{B}} = \mathcal{B}_{s}^{(\mathbf{u},\mathbf{y})}$; (ii) $(A,B)$ is controllable; (iii) $(C,A)$ is observable; and (iv) the signature matrix $\Sigma_{i}$ satisfies $A\Sigma_{i} = \Sigma_{i}A^{T}$, $B = \Sigma_{i}C^{T}$, and $D = D^{T}$. Then
\begin{align*}
\pi(\Sigma_{i}) &\geq \pi(\textnormal{Bez}(\hat{Q},\hat{P})), \\
\text{and } \nu(\Sigma_{i}) &\geq \nu(\textnormal{Bez}(\hat{Q},\hat{P})).
\end{align*} 
\end{lem}

Similarly, if $\mathcal{B}$ in (\ref{eq:bgd}) is controllable, then the following result follows from \cite{HugSmAI}.

\begin{lem}
\label{lem:rlctclb}
Let $\mathcal{B}$ in (\ref{eq:bgd}) be controllable. If $\mathcal{B}$ is the driving-point behavior of an RLCT network $N$, then $N$ contains at least $\pi(\textnormal{Bez}(Q,P))$ capacitors and at least $\nu(\textnormal{Bez}(Q,P))$ inductors.
\end{lem}

The purpose of this paper is to extend Lemmas \ref{thm:Gpbm} and \ref{lem:rlctclb} to include systems that need not be controllable. We will show that none of the lower bounds in those two lemmas are achievable when the corresponding behaviors are uncontrollable. Our main results are stated in the following two theorems. Here, for a given $\mathcal{B}$ of the form of (\ref{eq:bgd}), the notation $\zeta(Q,P)$ denotes the number of uncontrollable modes of $\mathcal{B}$ (see \ref{app:b}).
\begin{thm}
\label{thm:ssrsp}
Let $\hat{\mathcal{B}}$ be as in (\ref{eq:bgsqd}), and let $\mathcal{B}_{s}$ in (\ref{eq:bhssr}) be such that (i) $\hat{\mathcal{B}} = \mathcal{B}_{s}^{(\mathbf{u},\mathbf{y})}$; and (ii) the signature matrix $\Sigma_{i}$ satisfies $A\Sigma_{i} = \Sigma_{i}A^{T}$, $B = \Sigma_{i}C^{T}$, and $D = D^{T}$. Then 
\begin{align*}
\pi(\Sigma_{i}) &\geq \pi(\textnormal{Bez}(\hat{Q},\hat{P})) + \zeta(\hat{Q},\hat{P}), \\
\text{and } \nu(\Sigma_{i}) &\geq \nu(\textnormal{Bez}(\hat{Q},\hat{P})) + \zeta(\hat{Q},\hat{P}).
\end{align*} 
\end{thm}

\begin{thm}
\label{thm:brlctc}
If $\mathcal{B}$ in (\ref{eq:bgd}) is the driving-point behavior of an RLCT network $N$, then $N$ contains at least $\pi(\textnormal{Bez}(Q,P)) + \zeta(Q,P)$ capacitors and at least $\nu(\textnormal{Bez}(Q,P)) + \zeta(Q,P)$ inductors.
\end{thm}

These two theorems will be proven in Section \ref{sec:ram}, using results on the extended Cauchy index and McMillan degree of real-rational functions that are established in Section \ref{sec:ai}.

\begin{rem}
Using the results of Theorem \ref{thm:ssrsp}, it is easily verified that the realization in \cite[proof of Theorem 8]{THRB} is minimal, in the sense that the number of states with even and odd parity achieve the lower bounds established in Theorem \ref{thm:ssrsp}. Similarly, using the results in Theorem \ref{thm:brlctc}, it can be verified that the RLCT network realization in \cite[proof of Theorem 13]{THRB} is minimal, in the sense that the numbers of capacitors and inductors achieve the lower bounds in Theorem \ref{thm:brlctc}. However, we note that the lower bounds in Theorem \ref{thm:brlctc} do not apply to networks that contain gyrators in addition to resistors, inductors, capacitors and transformers (RLCTG networks). Specifically, if $\mathcal{B}$ in (\ref{eq:bgd}) is passive, then $\mathcal{B}$ can be realized as the driving-point behavior of an RLCTG network that contains exactly $\text{rank}(\textnormal{Bez}(Q,P)) + \zeta(Q,P)$ energy storage elements (inductors or capacitors). This is $\zeta(Q,P)$ fewer energy storage elements than are required to realize $\mathcal{B}$ with a reciprocal (i.e., RLCT) network.
\end{rem}

\section{The extended Cauchy index and the McMillan degree}
\label{sec:ai}
To prove Theorems \ref{thm:ssrsp} and \ref{thm:brlctc}, we will use the concepts of the \emph{McMillan degree} and \emph{extended Cauchy index} of a real-rational transfer function $H$, which we denote by $\delta(H)$ and $\gamma(H)$, respectively. The extended Cauchy index is defined for symmetric real-rational transfer functions as follows:
\begin{defn}\textnormal{\textbf{Extended Cauchy Index, \cite{HugSmAI}}}
\label{def:eci}
Let $H \in \mathbb{R}^{n \times n}(\xi)$ be symmetric. The extended Cauchy index of $H$ (denoted $\gamma(H)$) is the difference between the number of jumps in the eigenvalues of $H(\xi)$ from $-\infty$ to $+\infty$ less the number of jumps in the eigenvalues of $H(\xi)$ from $+\infty$ to $-\infty$ as $\xi$ increases from a point $a$ through $+\infty$ and then from $-\infty$ to $a$ again, for any $a \in \mathbb{R}$ that is not a pole of $H$.
\end{defn}
The extended Cauchy index and McMillan degree can be computed using matrix Bezoutians as follows:
\begin{lem}
\label{lem:bezcir}
Let $P, Q \in \mathbb{R}^{n \times n}[\xi]$, and let $H = Q^{-1}P$ be symmetric. Then 
\begin{align*}
\delta(H) = \pi(\textnormal{Bez}(Q,P)) + \nu(\textnormal{Bez}(Q,P)) = \text{rank}(\textnormal{Bez}(Q,P)),&\\
\text{and } \gamma(H) = \pi(\textnormal{Bez}(Q,P)) - \nu(\textnormal{Bez}(Q,P)).&
\end{align*}
\end{lem}

\begin{pf}
See \cite[Section 9]{HugSmAI}.
\end{pf}

We now state the main result in this section.
\begin{thm}
\label{thm:pnsr}
Let $H \in \mathbb{R}^{m \times m}(\xi)$ be symmetric, and let $S \in \mathbb{R}^{m \times n}$. Then
\begin{align}
\hspace*{-0.4cm} \gamma(H){+}\delta(H) &\geq \gamma(S^{T}HS) {-} \delta(S^{T}HS) {+} 2\delta(S^{T}H),\label{eq:psr1} \\
\hspace*{-0.4cm} \text{and } \gamma(H){-}\delta(H) &\leq \gamma(S^{T}HS) {+} \delta(S^{T}HS) {-} 2\delta(S^{T}H).\label{eq:psr2}
\end{align}
\end{thm} 

The proof of Theorem \ref{thm:pnsr} will use the following generalization of Sylvester's law of inertia.
\begin{lem}
\label{lem:sliwlk}
Let $P \in \mathbb{R}^{m \times m}$ be symmetric, and let $S \in \mathbb{R}^{m \times n}$. Then
\begin{align}
\pi(P) - \text{nullity}(S^{T}) &\leq \pi(S^{T}PS) \leq \pi(P), \\
\text{and } \nu(P) - \text{nullity}(S^{T}) &\leq \nu(S^{T}PS) \leq \nu(P).
\end{align}
\end{lem}

\begin{pf}
This result was shown in \cite{JDSLI}. We present an independent proof here that uses only basic linear algebra and the classical Sylvester's law of inertia. We prove the result for three different cases of increasing generality.

{\bf Case (a): $\boldsymbol{P = \text{diag}\begin{pmatrix}I_{n_{1}}& -I_{n_{2}}\end{pmatrix}}$ and $\boldsymbol{S}$ has full column rank.} \hspace{0.3cm} Let $S = \text{col}\begin{pmatrix}S_{1}& S_{2}\end{pmatrix}$ be partitioned compatibly with $P = \text{diag}\begin{pmatrix}I_{n_{1}}& -I_{n_{2}}\end{pmatrix}$; let the columns of $Y_{2}$ be a basis for the null space of $S_{1}$; and let $Y := \begin{bmatrix}Y_{1}& Y_{2}\end{bmatrix} \in \mathbb{R}^{m \times m}$ be nonsingular. If $\mathbf{z}$ is a real vector satisfying $S_{2}Y_{2}\mathbf{z} = 0$, then $SY\text{col}(0 \hspace{0.15cm} \mathbf{z}) = 0$. Since $S$ has full column rank and $Y$ is nonsingular, then this implies that $\mathbf{z} = 0$. It follows that $(S_{2}Y_{2})^{T}(S_{2}Y_{2}) > 0$. Next, let $((S_{2}Y_{2})^{T}(S_{2}Y_{2}))^{-1}(S_{2}Y_{2})^{T} \eqqcolon (S_{2}Y_{2})^{+}$ (which is the Moore-Penrose pseudoinverse of $S_{2}Y_{2}$), and let
\begin{equation*}
Z := \begin{bmatrix}Y_{1}& Y_{2}\end{bmatrix}\begin{bmatrix}I& 0\\ -(S_{2}Y_{2})^{+}S_{2}Y_{1}& I\end{bmatrix}.
\end{equation*}
Then $Z$ is nonsingular, and
\begin{equation}
SZ = \begin{bmatrix}S_{1}(I-Y_{2}(S_{2}Y_{2})^{+}S_{2})Y_{1}& 0\\ (I-(S_{2}Y_{2})(S_{2}Y_{2})^{+})S_{2}Y_{1}& S_{2}Y_{2}\end{bmatrix}.
\end{equation}
Since $(S_{2}Y_{2})^{T}(I{-}(S_{2}Y_{2})(S_{2}Y_{2})^{+}) {=} 0$, then $(SZ)^{T}P(SZ) = \text{diag}\begin{pmatrix}X_{11}& -(S_{2}Y_{2})^{T}(S_{2}Y_{2})\end{pmatrix}$ for some real symmetric matrix $X_{11} \in \mathbb{R}^{\text{rank}(S_{1}) \times \text{rank}(S_{1})}$. Also, by Sylvester's law of inertia, $\pi(S^{T}PS) = \pi((SZ)^{T}P(SZ))$ and $\nu(S^{T}PS) = \nu((SZ)^{T}P(SZ))$. Since, in addition, $0 > -(S_{2}Y_{2})^{T}(S_{2}Y_{2}) \in \mathbb{R}^{\text{nullity}(S_{1}) \times \text{nullity}(S_{1})}$, and $\text{nullity}(S_{1}) = \text{\#col}(S) {-} \text{rank}(S_{1}) = \text{rank}(S) {-} \text{rank}(S_{1}) \geq \text{\#row}(S) - \text{\#row}(S_{1}) - \text{nullity}(S^{T}) + \text{nullity}(S_{1}^{T})$, then
\begin{align*}
\pi((SZ)^{T}P(SZ)) &\leq \text{rank}(X_{11}) \leq \text{rank}(S_{1}) \leq n_{1}, \text{ and} \\
\nu((SZ)^{T}P(SZ)) &\geq \text{\#col}((S_{2}Y_{2})^{T}S_{2}Y_{2}) \\
&\geq \text{\#row}(S) - \text{\#row}(S_{1}) - \text{nullity}(S^{T}) \\
&=  n_{2} - \text{nullity}(S^{T}).
\end{align*}
Hence, $\pi(S^{T}PS) {\leq} \pi(P)$ and $\nu(S^{T}PS) {\geq} \nu(P) {-} \text{nullity}(S^{T})$. An entirely similar argument then shows that $\nu(S^{T}PS) \leq \nu(P)$ and $\pi(S^{T}PS) \geq \pi(P) - \text{nullity}(S^{T})$, and completes the proof of case (a).

{\bf Case (b): $\boldsymbol{P}$ nonsingular.} \hspace{0.3cm} Let the columns of $X_{2}$ be a basis for the null space of $S$, and let $X = \begin{bmatrix}X_{1}& X_{2}\end{bmatrix} \in \mathbb{R}^{n \times n}$ be nonsingular, so $SX_{1}$ has full column rank. Then $\text{diag}\begin{pmatrix}X_{1}^{T}S^{T}PSX_{1}& 0\end{pmatrix} = X^{T}S^{T}PSX$, which implies that 
\begin{align*}
\pi(S^{T}PS) &= \pi(X^{T}S^{T}PSX) = \pi(X_{1}^{T}S^{T}PSX_{1}), \text{ and}\\
\nu(S^{T}PS) &= \nu(X^{T}S^{T}PSX) = \nu(X_{1}^{T}S^{T}PSX_{1}).
\end{align*}
By Sylvester's law of inertia, there exists a nonsingular $R \in \mathbb{R}^{m \times m}$ such that $P = R^{T}\text{diag}\begin{pmatrix}I_{\pi(P)}& -I_{\nu(P)}\end{pmatrix}R$, so 
\begin{equation*}
X_{1}^{T}S^{T}PSX_{1} = (RSX_{1})^{T}\text{diag}\begin{pmatrix}I_{\pi(P)}& -I_{\nu(P)}\end{pmatrix}(RSX_{1}).
\end{equation*}
Since, in addition, $RSX_{1}$ has full column rank, then 
\begin{align*}
\pi(P) - \text{nullity}((RSX_{1})^{T}) &\leq \pi(S^{T}PS) \leq \pi(P), \text{ and}\\
\nu(P) - \text{nullity}((RSX_{1})^{T}) &\leq \nu(S^{T}PS) \leq \nu(P),
\end{align*}
by case (a). But $R$ and $X$ are nonsingular, and $RSX = \begin{bmatrix}RSX_{1}& 0\end{bmatrix}$, so $\text{nullity}((RSX_{1})^{T}) = \text{nullity}((RSX)^{T}) = \text{nullity}(S^{T})$. This proves case (b).

{\bf Case (c): general case.} \hspace{0.3cm} By Sylvester's law of inertia, there exists a nonsingular $R \in \mathbb{R}^{m \times m}$ such that $P = R^{T}\text{diag}\begin{pmatrix}I_{\pi(P)}& -I_{\nu(P)}& 0\end{pmatrix}R$. Let $R =: \text{col}\begin{pmatrix}R_{1}& R_{2}\end{pmatrix}$ with $R_{1} \in \mathbb{R}^{(\pi(P)+\nu(P)) \times m}$, so $(R_{1}S)^{T}\text{diag}\begin{pmatrix}I_{\pi(P)}& -I_{\nu(P)}\end{pmatrix}R_{1}S = S^{T}PS$. Thus,
\begin{align*}
\pi(P) - \text{nullity}((R_{1}S)^{T}) &\leq \pi(S^{T}PS) \leq \pi(P), \text{ and}\\
\nu(P) - \text{nullity}((R_{1}S)^{T}) &\leq  \nu(S^{T}PS) \leq \nu(P)
\end{align*} 
from case (b). But $R$ is nonsingular, so $R_{1}$ has full row rank, and so $\text{nullity}((R_{1}S)^{T}) \leq \text{nullity}(S^{T})$. This proves case (c).
\qed
\end{pf}

Prior to proving Theorem \ref{thm:pnsr}, we note that, if $H \in \mathbb{R}^{n \times n}(\xi)$ is proper, then the McMillan degree of $H$ may also be computed using Hankel matrices. The same is true of the extended Cauchy index (providing $H$ is also symmetric). Specifically, any $H \in \mathbb{R}_{p}^{n \times n}(\xi)$ is uniquely determined by its Markov parameters, which are the terms in the formal series expansion $H(\xi) = W_{-1} + W_{0}/\xi + W_{1}/ \xi^{2} + \ldots$. We let $\mathcal{H}_{r}(H)$ be the block matrix 
\begin{equation*}
\mathcal{H}_{r}(H) = \begin{bmatrix}W_{0}& W_{1}& \cdots & W_{r-1}\\ W_{1}& W_{2}& \cdots & W_{r}\\ \vdots& \vdots& &\vdots \\ W_{r-1}& W_{r}& \cdots & W_{2(r-1)}\end{bmatrix}, \hspace{0.2cm} (r = 1, 2, \ldots).
\end{equation*}
Then there exist unique integers $d$ and $N$ with $d \geq N$ such that $\text{rank}(\mathcal{H}_{r}(H)) = d$ for all $r \geq N$. The integer $d$ is equal to $\delta(H)$. Also, if $H$ is symmetric, then $\gamma(H) = \pi(\mathcal{H}_{r}(H)) - \nu(\mathcal{H}_{r}(H))$ for all $r \geq d$ \cite[Section 9]{HugSmAI}.

{\bf PROOF OF THEOREM \ref{thm:pnsr}}
We prove this first for the case that $H$ is proper, and then for the general case.

{\bf Case (a): $H$ proper.} \hspace{0.3cm} First, let $n = \delta(H)$, so $\delta(S^{T}H) \leq n$ and $\delta(S^{T}HS) \leq n$ by note \ref{nl:md4}. Also, let  $\hat{S}$ be the block diagonal matrix containing exactly $n$ diagonal blocks all equal to $S$, i.e., $\hat{S} := \text{diag}(S \hspace{0.15cm} \cdots \hspace{0.15cm} S)$. Then, note that $\mathcal{H}_{n}(S^{T}H) = \hat{S}^{T}\mathcal{H}_{n}(H)$ and $\mathcal{H}_{n}(S^{T}HS) = \hat{S}^{T}\mathcal{H}_{n}(H)\hat{S}$. It follows that 
\begin{align*}
\delta(H) + \gamma(H) &= 2\pi(\mathcal{H}_{n}(H)), \\
\delta(H) - \gamma(H) &= 2\nu(\mathcal{H}_{n}(H)), \\
\delta(S^{T}HS) + \gamma(S^{T}HS) &= 2\pi(\hat{S}^{T}\mathcal{H}_{n}(H)\hat{S}), \\
\delta(S^{T}HS) - \gamma(S^{T}HS) &= 2\nu(\hat{S}^{T}\mathcal{H}_{n}(H)\hat{S}), \displaybreak[3] \\
\delta(S^{T}H) &= \text{rank}(\hat{S}^{T}\mathcal{H}_{n}(H)), \\
\delta(H) &= \text{rank}(\mathcal{H}_{n}(H)), \\
\text{and } \delta(S^{T}HS) &= \text{rank}(\hat{S}^{T}\mathcal{H}_{n}(H)\hat{S}).
\end{align*}
It therefore suffices to show that 
\begin{align}
\hspace*{-0.4cm}\nu(\mathcal{H}_{n}(H)) &\geq  \text{rank}(\hat{S}^{T}\mathcal{H}_{n}(H)) - \pi(\hat{S}^{T}\mathcal{H}_{n}(H)\hat{S}), \text{ and}\label{nl:hmlbc1} \\
\hspace*{-0.4cm}\pi(\mathcal{H}_{n}(H)) &\geq \text{rank}(\hat{S}^{T}\mathcal{H}_{n}(H)) - \nu(\hat{S}^{T}\mathcal{H}_{n}(H)\hat{S}). \label{nl:hmlbc2}
\end{align}
To see (\ref{nl:hmlbc1}), let $n_{1} := \pi(\mathcal{H}_{n}(H))$ and $n_{2} := \nu(\mathcal{H}_{n}(H))$, so by Sylvester's law of inertia there exists a nonsingular real matrix $R$ such that $\mathcal{H}_{n}(H) = R^{T}\text{diag}(I_{n_{1}} \hspace{0.15cm} {-}I_{n_{2}} \hspace{0.15cm} 0)R$. Thus, the matrix $X$ formed from the first $n_{1}+n_{2}$ rows of $R$ has full row rank and satisfies $\mathcal{H}_{n}(H) = X^{T}\Sigma X$ where $\Sigma := \text{diag}(I_{n_{1}} \hspace{0.15cm} {-}I_{n_{2}})$. It follows that 
\begin{align*}
\text{nullity}((X\hat{S})^{T}) &= \text{\#row}(X\hat{S}) - \text{rank}(X\hat{S}) \\
&= n_{1} + n_{2} - \text{rank}(\hat{S}^{T}X^{T}), \\
\text{where } \text{rank}(\hat{S}^{T}X^{T}) &= \text{rank}(\begin{bmatrix}\hat{S}^{T}X^{T}\text{diag}\begin{pmatrix}I_{n_{1}}& \hspace*{-0.2cm} -I_{n_{2}}\end{pmatrix}& 0\end{bmatrix}) \\
&= \text{rank}(\hat{S}^{T}\mathcal{H}_{n}(H)R^{-1}) \\ &= \text{rank}(\hat{S}^{T}\mathcal{H}_{n}(H)).
\end{align*}
But $\hat{S}^{T}\mathcal{H}_{n}(H)\hat{S} = (X\hat{S})^{T}\Sigma(X\hat{S})$, so $\pi(\hat{S}^{T}\mathcal{H}_{n}(H)\hat{S}) \geq n_{1} - \text{nullity}((X\hat{S})^{T})$ from Lemma \ref{lem:sliwlk}, whereupon we obtain (\ref{nl:hmlbc1}). A similar argument then proves (\ref{nl:hmlbc2}).

{\bf Case (b): General case.} \hspace{0.3cm} This can be shown using a M\"{o}bius transformation to convert to the case with $H$ proper, in the manner of \cite[p.\ 224]{HugSmAI}. \qed

\section{Reciprocity and minimality, proof of Theorems \ref{thm:ssrsp} and \ref{thm:brlctc}}
\label{sec:ram}

{\bf PROOF OF THEOREM \ref{thm:ssrsp}}  (see p.\ \pageref{thm:ssrsp}). Let $\mathcal{A}(\xi) = \xi I - A$, and first note that 
\begin{equation*}
\hat{Q}^{-1}\hat{P} = D + C\mathcal{A}^{-1}B = D + (C\Sigma_{i})(\mathcal{A}\Sigma_{i})^{-1}B.
\end{equation*}
With the notation
\begin{equation*}
H := (\mathcal{A}\Sigma_{i})^{-1} \text{ and } S := B,
\end{equation*}
then $H$ is symmetric and $\hat{Q}^{-1}\hat{P} = D + S^{T}HS$. It then follows that $\gamma(\hat{Q}^{-1}\hat{P}) = \gamma(S^{T}HS)$ and $\delta(\hat{Q}^{-1}\hat{P}) = \delta(S^{T}HS)$ by note \ref{nl:md3} and \cite[Lemma 13]{HugSmAI}, so
\begin{align*}
\pi(\textnormal{Bez}(\hat{Q}, \hat{P})) &= \tfrac{1}{2}(\delta(S^{T}HS) + \gamma(S^{T}HS)), \text{ and} \\
\nu(\textnormal{Bez}(\hat{Q}, \hat{P})) &= \tfrac{1}{2}(\delta(S^{T}HS) - \gamma(S^{T}HS)).
\end{align*}
Next, note that $H = M^{-1}N$ where $M = \mathcal{A}\Sigma_{i}$ and $N = I$. Since, in addition, $A\Sigma_{i} = \Sigma_{i}A^{T}$, then from Definition \ref{def:bezd} we obtain $\textnormal{Bez}(M,N) = \Sigma_{i}$. Thus, $\pi(\Sigma_{i}) = \tfrac{1}{2}(\delta(H) + \gamma(H))$ and $\nu(\Sigma_{i}) = \tfrac{1}{2}(\delta(H) - \gamma(H))$. Then, from Lemma \ref{lem:bezcir} and Theorem \ref{thm:pnsr},
\begin{align*}
\pi(\Sigma_{i}) &\geq \pi(\textnormal{Bez}(\hat{Q}, \hat{P})) + \delta(S^{T}H) - \delta(S^{T}HS), \text{ and}\\
\nu(\Sigma_{i}) &\geq \nu(\textnormal{Bez}(\hat{Q},\hat{P})) + \delta(S^{T}H) - \delta(S^{T}HS).
\end{align*}
But $S^{T}H = C\mathcal{A}^{-1}$ and $\delta(S^{T}HS) = \delta(\hat{Q}^{-1}\hat{P})$, so by note \ref{nl:md2} it remains to show that $\delta(C\mathcal{A}^{-1}) = \Delta([{-}\hat{P} \hspace{0.2cm} \hat{Q}])$. To see this, we note initially from \cite[Lemma 1]{THRB} that there exist polynomial matrices $U, V, Y, Z, E, F$, and $G$ such that
\begin{equation}
\begin{bmatrix}Y& Z\\ U& V\end{bmatrix}\begin{bmatrix}-D& I& -C\\ -B& 0& \mathcal{A}\end{bmatrix} = \begin{bmatrix}-\hat{P}& \hat{Q}& 0\\ -E& -F& G\end{bmatrix}, \label{eq:ssrut}
\end{equation} 
in which $G$ is nonsingular and the leftmost matrix is unimodular. By pre-multiplying both sides of this equation by the inverse of the leftmost matrix, and comparing the rightmost block column in the resulting equation, we obtain a relationship of the form
\begin{equation*}
\begin{bmatrix}W_{1} \\ W_{2}\end{bmatrix}G =  \begin{bmatrix}-C \\ \mathcal{A}\end{bmatrix},
\end{equation*}
in which $\text{col}(W_{1} \hspace{0.15cm} W_{2})(\lambda)$ has full column rank for all $\lambda \in \mathbb{C}$. In particular, since $\mathcal{A}$ is nonsingular, then so too is $W_{2}$, and $C\mathcal{A}^{-1} = W_{1}W_{2}^{-1}$ which is proper. It follows from note \ref{nl:md2} that $\text{\#col}(A) = \Delta([{-}C^{T} \hspace{0.15cm} \mathcal{A}^{T}]) = \deg{(\det{(G)})} + \delta(W_{1}W_{2}^{-1}) = \deg{(\det{(G)})} + \delta(C\mathcal{A}^{-1})$. Also, from (\ref{eq:ssrut}),
\begin{align*}
\Delta([{-}\hat{P} \hspace{0.2cm} \hat{Q}]) + \deg{(\det{(G)})} &= \Delta\left(\begin{bmatrix}-D& I& -C\\ -B& 0& \mathcal{A}\end{bmatrix}\right) \\
&= \text{\#col}(A),
\end{align*}
whereupon we conclude that $\Delta([{-}\hat{P} \hspace{0.2cm} \hat{Q}]) = \delta(C\mathcal{A}^{-1})$. \qed

{\bf PROOF OF THEOREM \ref{thm:brlctc}}  (see p.\ \pageref{thm:brlctc}). We let $\tilde{\mathcal{B}}$ denote the behavior of $N$, and we will show that there exist compatible partitions of the driving-point currents and voltages as $(\mathbf{i}_{a}, \mathbf{i}_{b})$ and $(\mathbf{v}_{a}, \mathbf{v}_{b})$; compatible partitions of the inductor currents and voltages as $(\mathbf{i}_{La}, \mathbf{i}_{Lb})$ and $(\mathbf{v}_{La}, \mathbf{v}_{Lb})$; compatible partitions of the capacitor currents and voltages as $(\mathbf{i}_{Ca}, \mathbf{i}_{Cb})$ and $(\mathbf{v}_{Ca}, \mathbf{v}_{Cb})$; a state-space model $\mathcal{B}_{s}$ as in (\ref{eq:bhssr}); and a signature matrix $\Sigma_{i} := \text{diag}({-}I \hspace{0.15cm} I)$ in which the number of $-1$ (resp., $+1$) entries is equal to the number of entries in $\mathbf{i}_{La}$ (resp., $\mathbf{v}_{Ca}$); such that $A\Sigma_{i} = \Sigma_{i}A^{T}, B = \Sigma_{i}C^{T}, D = D^{T}$, and $\tilde{\mathcal{B}}^{(\text{col}(\mathbf{i}_{a} \hspace{0.15cm} {-}\mathbf{v}_{b}), \text{col}(\mathbf{v}_{a} \hspace{0.15cm} \mathbf{i}_{b}))} \eqqcolon \hat{\mathcal{B}} = \mathcal{B}_{s}^{(\mathbf{u},\mathbf{y})}$. But it is easily shown that $\hat{\mathcal{B}}$ also takes the form of (\ref{eq:bgsqd}), where (i) $\hat{Q}(z)\hat{P}(w)^{T}-\hat{P}(z)\hat{Q}(w)^{T} = Q(z)P(w)^{T}-P(z)Q(w)^{T}$, so $\textnormal{Bez}(Q,P) = \textnormal{Bez}(\hat{Q},\hat{P})$ by Definition \ref{def:bezd}; and (ii) there exists a nonsingular $S \in \mathbb{R}^{2n\times 2n}$ such that $[-\hat{P} \hspace{0.15cm} \hat{Q}] = [-P \hspace{0.15cm} Q]S$, so it is easily shown from notes \ref{nl:b1}--\ref{nl:b2} that $\zeta(\hat{Q},\hat{P}) = \zeta(Q,P)$. The present theorem then follows from Theorem \ref{thm:ssrsp}.

To see that $\hat{\mathcal{B}}$ has a state-space representation of the form indicated in the previous paragraph, we first note from \cite[Theorem 5]{HUGIFAC} that there exist partitions of the driving-point, inductor and capacitor currents and voltages as in that paragraph such that, with the notation
\begin{align*}
\mathbf{e} &{=} \text{col}(\mathbf{i}_{a} \hspace{0.15cm} \mathbf{v}_{b}), \mathbf{r} {=} \text{col}(\mathbf{v}_{a} \hspace{0.15cm} \mathbf{i}_{b}), \mathbf{e}_{1a} {=} \text{col}(\mathbf{i}_{La} \hspace{0.15cm} \mathbf{v}_{Ca}),\\
\mathbf{r}_{1a} &{=} \text{col}(\mathbf{v}_{La} \hspace{0.15cm} \mathbf{i}_{Ca}), \mathbf{e}_{1b} {=} \text{col}(\mathbf{v}_{Lb} \hspace{0.15cm} \mathbf{i}_{Cb}) \text{ and } \mathbf{r}_{1b} {=} \text{col}(\mathbf{i}_{Lb} \hspace{0.15cm} \mathbf{v}_{Cb});
\end{align*}
with $\Sigma_{e}, \Sigma_{1}$ and $\Sigma_{2}$ signature matrices that, partitioned compatibly with $\mathbf{e}, \mathbf{e}_{1a}$ and $\mathbf{e}_{1b}$, respectively, take the form 
\begin{equation*}
\Sigma_{e} {=} \text{diag}(I \hspace{0.15cm} {-}I), \Sigma_{1} {=} \text{diag}({-}I \hspace{0.15cm} I), \text{ and }\Sigma_{2} {=} \text{diag}({-}I \hspace{0.15cm} I);
\end{equation*}
and with $\Lambda_{1}$ (resp., $\Lambda_{2}$) the diagonal matrix whose entries correspond to the inductances and capacitances relating to the corresponding entry in $\mathbf{e}_{1a}$ (resp., $\mathbf{e}_{1b}$); then $\tilde{\mathcal{B}}^{(\mathbf{e}, \mathbf{r}, \mathbf{e}_{1a}, \mathbf{r}_{1a}, \mathbf{e}_{1b}, \mathbf{r}_{1b})}$ is determined by equations of the form:
\begin{align}
&\hspace*{-0.5cm}\mathbf{e}_{2a} {=} \Sigma_{1}\mathbf{e}_{1a}, \hspace{0.05cm} \mathbf{r}_{2a} {=} {-}\Sigma_{1}\mathbf{r}_{1a}, \hspace{0.05cm} \mathbf{r}_{2b} {=} \Sigma_{2}\mathbf{r}_{1b} \text{ and } \mathbf{e}_{2b} {=} {-}\Sigma_{2}\mathbf{e}_{1b},\label{eq:bic} \\
&\hspace*{-0.5cm}\mathbf{r}_{1a} = \Lambda_{1}\tfrac{d\mathbf{e}_{1a}}{dt}, \hspace{0.15cm} \mathbf{e}_{1b} = \Lambda_{2}\tfrac{d\mathbf{r}_{1b}}{dt},\label{eq:dpblcb} \\
&\hspace*{-0.5cm}\text{and }\begin{bmatrix}M_{11}& M_{12}& 0\\M_{21}& M_{22}& M_{23}\\ 0& -M_{23}^{T}& 0\end{bmatrix}\begin{bmatrix}\mathbf{e} \\ \mathbf{e}_{2a}\\ \mathbf{e}_{2b}\end{bmatrix} = \begin{bmatrix}\mathbf{r} \\ \mathbf{r}_{2a}\\ \mathbf{r}_{2b}\end{bmatrix}.\label{eq:dpbrtge}
\end{align}

Here, (\ref{eq:dpbrtge}) describes the driving-point behavior of a network containing only resistors and transformers, so
\begin{equation*}
\begin{bmatrix}M_{11}& M_{12}& 0\\M_{21}& M_{22}& M_{23}\\ 0& -M_{23}^{T}& 0\end{bmatrix}\begin{bmatrix}\Sigma_{e}& 0& 0\\ 0& -\Sigma_{1}& 0\\ 0& 0& \Sigma_{2}\end{bmatrix}
\end{equation*}
is symmetric by \cite[Theorem 2.8.1]{AndVong}. In particular,
\begin{align}
&M_{11}\Sigma_{e} \text{ and } M_{22}\Sigma_{1} \text{ are symmetric}, \nonumber \\
&M_{21}\Sigma_{e} = -\Sigma_{1}M_{12}^{T}, \text{ and} \nonumber \\
&M_{23}\Sigma_{2} = \Sigma_{1}M_{23}.\label{eq:sr}
\end{align}
Moreover, $\Omega := \Lambda_{1} + \Sigma_{1}M_{23}\Sigma_{2}\Lambda_{2}\Sigma_{2}M_{23}^{T}\Sigma_{1}$ satisfies $\Omega > 0$ and $\Omega\Sigma_{1} = \Lambda_{1}\Sigma_{1} + M_{23}\Lambda_{2}\Sigma_{2}M_{23}^{T}$, which is symmetric. In particular, it follows that $\Omega$ (partitioned compatibly with $\Sigma_{1}$) takes the form $\Omega = \text{diag}(\Omega_{11} \hspace{0.15cm} \Omega_{22})$ where $\Omega_{11}, \Omega_{22} >0$. Next, from \cite[proof of Theorem 5]{HUGIFAC}, $\tilde{\mathcal{B}}^{(\Sigma_{e}\mathbf{e}, \mathbf{r}, \mathbf{e}_{1a})} = \tilde{\mathcal{B}}_{s}$ where $\tilde{\mathcal{B}}_{s} = \lbrace (\mathbf{u}, \mathbf{y}, \tilde{\mathbf{x}}) \in \mathcal{L}_{1}^{\text{loc}}\left(\mathbb{R}, \mathbb{R}^{n}\right) {\times} \mathcal{L}_{1}^{\text{loc}}\left(\mathbb{R}, \mathbb{R}^{n}\right) {\times} \mathcal{L}_{1}^{\text{loc}}\left(\mathbb{R}, \mathbb{R}^{d}\right) \mid \tfrac{d\tilde{\mathbf{x}}}{dt} = \tilde{A}\tilde{\mathbf{x}} + \tilde{B}\mathbf{u} \text{ and } \mathbf{y} = \tilde{C}\tilde{\mathbf{x}} + \tilde{D}\mathbf{u}\rbrace$, with
\begin{align*}
\tilde{A} &= -\Omega^{-1}\Sigma_{1}M_{22}\Sigma_{1}, \hspace{0.15cm} \tilde{B} = -\Omega^{-1}\Sigma_{1}M_{21}\Sigma_{e}, \\ 
\tilde{C} &= M_{12}\Sigma_{1} \text{ and } \tilde{D} = M_{11}\Sigma_{e}.
\end{align*}
Finally, we let $\Omega^{1/2}, \Omega_{11}^{1/2}$ and $\Omega_{22}^{1/2}$ denote the positive-definite square roots of $\Omega, \Omega_{11}$ and $\Omega_{22}$, respectively, so $\Omega^{1/2} = \text{diag}(\Omega_{11}^{1/2} \hspace{0.15cm} \Omega_{22}^{1/2})$, and $\Omega^{1/2}$ commutes with $\Sigma_{1}$. We then let $\mathcal{B}_{s}$ be as in (\ref{eq:bhssr}) with
\begin{equation*}
A := \Omega^{1/2}\tilde{A}\Omega^{-1/2}, \hspace{0.1cm} B:= \Omega^{1/2}\tilde{B}, \hspace{0.1cm} C:= \tilde{C}\Omega^{-1/2}, D := \tilde{D},
\end{equation*}
and it follows that $\mathcal{B}^{(\text{col}(\mathbf{i}_{a} \hspace{0.15cm} {-}\mathbf{v}_{b}), \text{col}(\mathbf{v}_{a} \hspace{0.15cm} \mathbf{i}_{b}))} \eqqcolon \hat{\mathcal{B}} = \tilde{\mathcal{B}}_{s}^{(\mathbf{u},\mathbf{y})} = \mathcal{B}_{s}^{(\mathbf{u},\mathbf{y})}$. Furthermore, it is straightforward to verify that 
\begin{align*}
A = -\Sigma_{1}\Omega^{-1/2}M_{22}\Sigma_{1}\Omega^{-1/2}, \hspace{0.1cm} B = -\Sigma_{1}\Omega^{-1/2}M_{21}\Sigma_{e}, \\
C = M_{12}\Sigma_{1}\Omega^{-1/2} \text{ and } D = M_{11}\Sigma_{e}
\end{align*}
Thus, with $\Sigma_{i} = \Sigma_{1}$, then it follows from (\ref{eq:sr}) that $A\Sigma_{i} = \Sigma_{i}A^{T}$, $B = \Sigma_{i}C^{T}$ and $D = D^{T}$. \qed 

\appendix
\section{Behaviors and state-space realizations}
\label{app:b}
In this appendix, we present a number of relevant results on behaviors and state-space realizations. For references, see \cite{JWIMTSC, PRSMLS, THBRSF}.
\begin{remunerate}
\labitem{A\arabic{muni}}{nl:b1} Let $\mathcal{B}$ be as in (\ref{eq:bgd}). Then, by \cite[Lemma 10]{THRB}, there exist $F, \tilde{P}, \tilde{Q}, U, V \in \mathbb{R}^{n \times n}[\xi]$ such that 
\begin{enumerate}[label=(\roman*), ref=(\roman*)]
\item $P = F\tilde{P}$ and $Q = F\tilde{Q}$; and\label{nl:cadc1i}
\item $\begin{bmatrix}\tilde{P}& \hspace{0.1cm}{-}\tilde{Q}\\ U& \hspace{0.1cm}V\end{bmatrix}$ is unimodular.\label{nl:cadc1ii}
\end{enumerate}
Also, if $F, \tilde{P}, \tilde{Q}, U, V \in \mathbb{R}^{n \times n}[\xi]$ satisfy \ref{nl:cadc1i}--\ref{nl:cadc1ii}, then 
$(\mathbf{i}, \mathbf{v}) \in \mathcal{B} \iff$ there exist $(\mathbf{i}_{1}, \mathbf{v}_{1}) \in \mathcal{B}_{c}$ and $(\mathbf{i}_{2}, \mathbf{v}_{2}) \in \mathcal{B}_{a}$ with $\mathbf{i} = \mathbf{i}_{1} {+} \mathbf{i}_{2}$ and $\mathbf{v} = \mathbf{v}_{1} {+} \mathbf{v}_{2}$, where \newline $\mathcal{B}_{c} := \lbrace (\mathbf{i}, \mathbf{v}) \in \mathcal{L}_{1}^{\text{loc}}\left(\mathbb{R}, \mathbb{R}^{n}\right) \times \mathcal{L}_{1}^{\text{loc}}\left(\mathbb{R}, \mathbb{R}^{n}\right) \mid \tilde{P}(\tfrac{d}{dt})\mathbf{i} = \tilde{Q}(\tfrac{d}{dt})\mathbf{v} \rbrace$ and \newline $\mathcal{B}_{a} := \lbrace (\mathbf{i}, \mathbf{v}) \in \mathcal{L}_{1}^{\text{loc}}\left(\mathbb{R}, \mathbb{R}^{n}\right) \times \mathcal{L}_{1}^{\text{loc}}\left(\mathbb{R}, \mathbb{R}^{n}\right) \mid P(\tfrac{d}{dt})\mathbf{i} = Q(\tfrac{d}{dt})\mathbf{v} \text{ and } U(\tfrac{d}{dt})\mathbf{i} = -V(\tfrac{d}{dt})\mathbf{v} \rbrace$.\label{nl:cadc1}
\labitem{A\arabic{muni}}{nl:b2} In note \ref{nl:b1}, we define $\zeta(Q,P):= \deg{(\det{(F)})}$, and it is easily shown that this is invariant of the specific choice of decomposition in that lemma. Here, $\zeta(Q,P)$ represents the number of uncontrollable modes of $\mathcal{B}$.  Specifically, for any given decomposition as in note \ref{nl:b1}, it is easily shown that $\mathcal{B}_{a}$ is a subspace of $\mathcal{L}_{1}^{\text{loc}}\left(\mathbb{R}, \mathbb{R}^{n}\right) \times \mathcal{L}_{1}^{\text{loc}}\left(\mathbb{R}, \mathbb{R}^{n}\right)$ of dimension $\zeta(Q,P)$ (see \cite[Theorem 3.2.16]{JWIMTSC}); and $\mathbf{i}_{1}$ and $\mathbf{i}_{2}$ (resp., $\mathbf{v}_{1}$ and $\mathbf{v}_{2}$) are uniquely determined by $\mathbf{i}$ (resp., $\mathbf{v}$). To see this, suppose that (a) $(\mathbf{i}_{1a}, \mathbf{v}_{1a}), (\mathbf{i}_{1b}, \mathbf{v}_{1b}) \in \mathcal{B}_{c}$; (b) $(\mathbf{i}_{2a}, \mathbf{v}_{2a}), (\mathbf{i}_{2b}, \mathbf{v}_{2b}) \in \mathcal{B}_{a}$; (c) $\mathbf{i}_{1a} + \mathbf{i}_{2a} = \mathbf{i}_{1b} + \mathbf{i}_{2b}$; and (d) $\mathbf{v}_{1a} + \mathbf{v}_{2a} = \mathbf{v}_{1b} + \mathbf{v}_{2b}$. Then $\tilde{P}(\tfrac{d}{dt})(\mathbf{i}_{1a}-\mathbf{i}_{1b}) - \tilde{Q}(\tfrac{d}{dt})(\mathbf{v}_{1a}-\mathbf{v}_{1b}) = 0$, so $\tilde{P}(\tfrac{d}{dt})(\mathbf{i}_{2a}-\mathbf{i}_{2b}) - \tilde{Q}(\tfrac{d}{dt})(\mathbf{v}_{2a}-\mathbf{v}_{2b}) = 0$ by (c)--(d). Since, in addition, $U(\tfrac{d}{dt})(\mathbf{i}_{2a}-\mathbf{i}_{2b}) + V(\tfrac{d}{dt})(\mathbf{v}_{2a}-\mathbf{v}_{2b}) = 0$, and condition \ref{nl:cadc1ii} of note \ref{nl:b1} holds, then $\mathbf{i}_{2a} = \mathbf{i}_{2b}$ and $\mathbf{v}_{2a} = \mathbf{v}_{2b}$ by \cite[Theorem 3.2.16]{JWIMTSC}, whence $\mathbf{i}_{1a} = \mathbf{i}_{1b}$ and $\mathbf{v}_{1a} = \mathbf{v}_{1b}$. 
\end{remunerate}

\section{McMillan degree}
\label{app:md}
Here, we provide several useful properties concerning the McMillan degree of a real-rational function.
\begin{remunerate}
\labitem{B\arabic{muni}}{nl:md1} Let $P,Q, \tilde{P}$ and $\tilde{Q}$ be as in note \ref{nl:b1} with $Q$ nonsingular. Then $Q^{-1}P = \tilde{Q}^{-1}\tilde{P}$, and if, in addition, $Q^{-1}P$ is proper, then $\delta(Q^{-1}P) {=} \deg{(\det{(\tilde{Q})})}$ \cite[Section 3]{BitAnd}.
\labitem{B\arabic{muni}}{nl:md2} Let $P$ and $Q$ be as in note \ref{nl:b1}. If $Q$ is nonsingular, then $Q^{-1}P$ is proper if and only if $\Delta([{-}P \hspace{0.25cm} Q]) = \deg{(\det{(Q)})}$ \cite[proof of Theorem 3.3.22]{JWIMTSC}. In particular, if $Q^{-1}P$ is proper and $\tilde{P}, \tilde{Q}$ and $F$ are as in note \ref{nl:b1}, then $\delta(Q^{-1}P) = \Delta([{-}\tilde{P} \hspace{0.25cm} \tilde{Q}])$ and $\Delta([{-}P \hspace{0.25cm} Q]) = \zeta(Q,P) + \delta(Q^{-1}P)$ (see notes \ref{nl:b1} and \ref{nl:md1}).
\labitem{B\arabic{muni}}{nl:md3} If $X \in \mathbb{R}^{m \times n}$ and $Y \in \mathbb{R}^{m \times n}(\xi)$, then $\delta(X + Y) = \delta (Y)$ \cite[Chapter 3]{AndVong}.
\labitem{B\arabic{muni}}{nl:md4} If $S \in \mathbb{R}^{k \times l}, T \in \mathbb{R}^{m \times n}$ and $Y \in \mathbb{R}^{l \times m}(\xi)$, then $\delta(SYT) \leq \delta(Y)$ \cite[Chapter 3]{AndVong}.
\end{remunerate}
\bibliographystyle{elsarticle-num}        
\bibliography{recip}

\end{document}